\documentclass[aps,prb,a4paper,10pt,twocolumn,showpacs,floatfix,superscriptaddress,amsmath,amsfonts,amssymb,preprintnumbers,longbibliography]{revtex4-1}
\usepackage{float}

\usepackage{dcolumn,graphicx,color,booktabs,microtype,afterpage}
\graphicspath{{./}{figure/}}
\usepackage[charter,greekuppercase=italicized]{mathdesign}
\usepackage{sidecap}
\renewcommand{\tablename}{Table}
\makeatletter\renewcommand{\fnum@figure}[1]{\figurename~\thefigure.~}\makeatother
\makeatletter\renewcommand{\fnum@table}[1]{\tablename~\thetable.}\makeatother

\newcount\hh \newcount\mm
\hh=\time \divide\hh by 60
\mm=\hh \multiply\mm by 60 \mm=-\mm
\advance\mm by \time
\def\now{\number\hh:\ifnum\mm<10{}0\fi\number\mm}

\usepackage[colorlinks,plainpages=false,linkcolor=blue,urlcolor=blue,citecolor=blue,pdfpagemode=UseNone,pdfstartview=FitBH]{hyperref}

\begin{document}

\makeatletter\renewcommand{\ps@plain}{%
\def\@evenhead{\hfill\itshape\rightmark}%
\def\@oddhead{\itshape\leftmark\hfill}%
\renewcommand{\@evenfoot}{\hfill\small{--~\thepage~--}\hfill}%
\renewcommand{\@oddfoot}{\hfill\small{--~\thepage~--}\hfill}%
}\makeatother\pagestyle{plain}

\preprint{\textit{Preprint: \today, \now.}} 

\title{Strong- to weak-coupling superconductivity in high-$T_c$ bismuthates: \\ 
       revisiting the phase diagram via $\mu$SR}

%
\author{T.\ Shang}\email[Corresponding authors:\\]{tian.shang@psi.ch}
\affiliation{Laboratory for Multiscale Materials Experiments, Paul Scherrer Institut, Villigen CH-5232, Switzerland}
\affiliation{Physik-Institut, Universit\"{a}t Z\"{u}rich, Winterthurerstrasse 190, CH-8057 Z\"{u}rich, Switzerland}
%
\author{D.\ J.\ Gawryluk}
\email{dariusz.gawryluk@psi.ch}\thanks{On leave from Institute of Physics, Polish Academy of Sciences, Aleja Lotnikow 32/46, PL-02-668 Warsaw, Poland.}
\affiliation{Laboratory for Multiscale Materials Experiments, Paul Scherrer Institut, Villigen CH-5232, Switzerland}
\author{M.\ Naamneh}
\affiliation{Swiss Light Source, Paul Scherrer Institut, Villigen CH-5232, Switzerland}
\affiliation{Department of Physics, Ben-Gurion University of the Negev, Beer-Sheva, 84105, Israel}
\author{Z.\ Salman}
\affiliation{Laboratory for Muon-Spin Spectroscopy, Paul Scherrer Institut, CH-5232 Villigen PSI, Switzerland}
\author{Z.~Guguchia}
\affiliation{Laboratory for Muon-Spin Spectroscopy, Paul Scherrer Institut, CH-5232 Villigen PSI, Switzerland}
\author{M.\ Medarde}
\affiliation{Laboratory for Multiscale Materials Experiments, Paul Scherrer Institut, Villigen CH-5232, Switzerland}
\author{M.\ Shi}
\affiliation{Swiss Light Source, Paul Scherrer Institut, Villigen CH-5232, Switzerland}

\author{N.\ C.\ Plumb}
\affiliation{Swiss Light Source, Paul Scherrer Institut, Villigen CH-5232, Switzerland}
\author{T.\ Shiroka}
\affiliation{Laboratory for Muon-Spin Spectroscopy, Paul Scherrer Institut, CH-5232 Villigen PSI, Switzerland}
\affiliation{Laboratorium f\"ur Festk\"orperphysik, ETH Z\"urich, CH-8093 Zurich, Switzerland}
\begin{abstract}
Several decades after the discovery of 
superconductivity 
in bismuthates, the strength of their electron-phonon coupling and its 
evolution  with doping remain puzzling. 
To clarify these issues, polycrystalline hole-doped Ba$_{1-x}$K$_{x}$BiO$_3$ ($0.1 \le x \le 0.6$) 
samples were systematically synthesized and their bulk- 
and microscopic superconducting properties were investigated by means of 
magnetic susceptibility and muon-spin rotation/relaxation ($\mu$SR), respectively.
The phase diagram of Ba$_{1-x}$K$_{x}$BiO$_3$ was reliably  
extended up to $x = 0.6$, which is still found to be a bulk superconductor. 
The lattice parameter $a$ increases linearly with K-content, 
implying a homogeneous chemical doping. 
The low-temperature superfluid density, measured via transverse-field 
(TF)-$\mu$SR, indicates an isotropic fully-gapped superconducting state 
with zero-temperature gaps $\Delta_0/k_\mathrm{B}T_c = 2.15$, 2.10, and 1.75, and 
magnetic penetration depths $\lambda_0 = 219$, 184, and 279\,nm for 
$x = 0.3$, 0.4, and 0.6, respectively. 
A change in the superconducting gap, from a nearly ideal 
BCS value
(1.76\,$k_\mathrm{B}$$T_c$ in the weak coupling case) 
in the overdoped $x=0.6$ region, to much higher values in the 
optimally-doped case, implies a gradual decrease in electron-phonon 
coupling with doping.
\end{abstract}



\maketitle\enlargethispage{3pt}

\vspace{-5pt}
\section{Introduction}\enlargethispage{8pt}
%
Decades ago, 
superconductivity (SC) with critical temperatures $T_c$ up to 34\,K was discovered in perovskite-type bismuthates~\cite{Cava1988,Mattheiss1988,Sleight2015}. 
Despite extensive studies using various techniques, their pairing mechanism 
is still under debate~\cite{Sleight2015}. The parent compound BaBiO$_3$ is an insulator, 
which exhibits charge-density-wave (CDW) order and undergoes multiple structural phase transitions~\cite{Munakata1992,Sato1989,Plumb2016}.
The suppression of the insulating character and of the CDW order upon 
Ba/K or Bi/Pb substitutions, eventually leads to a superconducting phase 
in Ba$_{1-x}$K$_{x}$BiO$_3$ or BaBi$_{1-y}$Pb$_y$O$_3$~\cite{Sleight2015}, 
with the highest $T_c$ reaching 34\,K near $x_\mathrm{K} \sim 0.37$  
[see Fig.~\ref{fig:phase_diagram}(c)].
Neutron powder diffraction measurements show  
that, for $0.2 \le x \le 0.37$, Ba$_{1-x}$K$_{x}$BiO$_3$ exhibits a cubic-to-orthorhombic phase transition before entering the superconducting phase, while for $x \ge 0.37$, the cubic structure persist down to the superconducting phase~\cite{Pei1990,Hinks1988}.
However, close to $x = 0.32$, 
a mixture of different phases has also been found~\cite{Hinks1988}.
According to our Rietveld refinements of x-ray powder diffraction 
(XRD) data, samples with $0.1 \leq x \leq 0.25$ exhibit a mixture of 
orthorhombic and cubic phases~\cite{Supple}.
Later on, it was confirmed that, at low temperatures, the distortion 
observed in the superconducting Ba$_{1-x}$K$_{x}$BiO$_3$ samples with 
$x \sim 0.32$--0.4 is more consistent with a tetragonal symmetry~\cite{Braden2000}.
Systematic studies of the Ba$_{1-x}$K$_{x}$BiO$_3$ crystal 
structure as a function of doping and temperature are clearly 
not only in high demand, but also crucial to understand their properties.
Indeed, bismuthates rank among the most interesting systems, where the interplay between 
structural-, charge-,  and electronic instabilities gives rise to new and remarkable phenomena.

According to the phase diagram shown in Fig.~\ref{fig:phase_diagram}, 
the lack of either magnetic order or magnetic fluctuations in bismuthates hints at a 
nonmagnetic superconducting pairing. 
To date, two pairing mechanisms, in the two extremes of coupling strengths, have been proposed in order to explain the unexpectedly high $T_{c}$ 
of doped bismuthates. One mechanism considers a strong coupling of electrons to high-energy phonon modes, leading to the formation of \emph{polarons.} 
The polarons then bind into Cooper pairs through a retarded electron-phonon interaction with the low-energy phonon modes~\cite{Rice1981,Alexandrov1996,Zhao2001}. Recently, a large electron-phonon coupling constant $\lambda_\mathrm{ep}$ $>$ 1, strong enough to account for the high $T_c$ in Ba$_{1-x}$K$_{x}$BiO$_3$, has been experimentally and theoretically proposed~\cite{Wen2018,Li2019}.
An alternative mechanism suggests that the pairing is  
mediated by high-energy \emph{charge excitations}~\cite{Batlogg1988}. 
This mechanism does not require a strong coupling, i.e., the superconducting 
energy gap can be close to 1.76\,$k_\mathrm{B}$$T_c$, the canonical 
weak-coupling BCS-theory value.\\
In case of strong coupling, the energy gap is much larger than the BCS value 
(for instance, for $x = 0.37$, $\Delta_{0}$ is circa 2.2\,$k_\mathrm{B}$$T_c$~\cite{Zhao2007}). Upon increasing the K-content, the bond/charge disproportionation weakens and eventually it disappears. Nevertheless, the electron pairs may still survive and condense, leading to a superconducting phase, albeit with a reduced $T_{c}$ value. 
One could make an educated guess and suggest a \emph{doping-dependent coupling strength}, 
a scenario which does not exclude either of the above mechanisms, 
thus accounting for the widely different experimental results. 

To validate the above hypothesis on
the evolution of the coupling strength with doping, the study of the superconducting gap 
and of its symmetry across the entire  
phase diagram of Ba$_{1-x}$K$_{x}$BiO$_3$ is crucial, in particular 
in the over-doped region, mostly overlooked due to the lack of high-quality samples. 
Here, by improving the synthesis conditions, we could obtain high-quality Ba$_{1-x}$K$_{x}$BiO$_3$ samples 
($0.1 \le x \le 0.6$), of which those with $x = 0.3$--0.6 show bulk superconductivity.
In this paper, we report on the systematic magnetization- and $\mu$SR 
investigation of the hole-doped Ba$_{1-x}$K$_{x}$BiO$_3$ system in the range 
$0.1 \le x \le 0.6$. 
By using transverse-field (TF-) $\mu$SR measurements, we study the microscopic 
superconducting properties, including the gap symmetry, the zero-temperature 
magnetic penetration depth, and the gap values across the whole 
superconducting phase region of Ba$_{1-x}$K$_{x}$BiO$_3$, to clearly 
demonstrate the decrease of the SC-coupling strength with doping. 

\section{Experimental details\label{sec:details}}
\begin{figure}[ht]
	\centering
	\includegraphics[width=0.48\textwidth,angle=0]{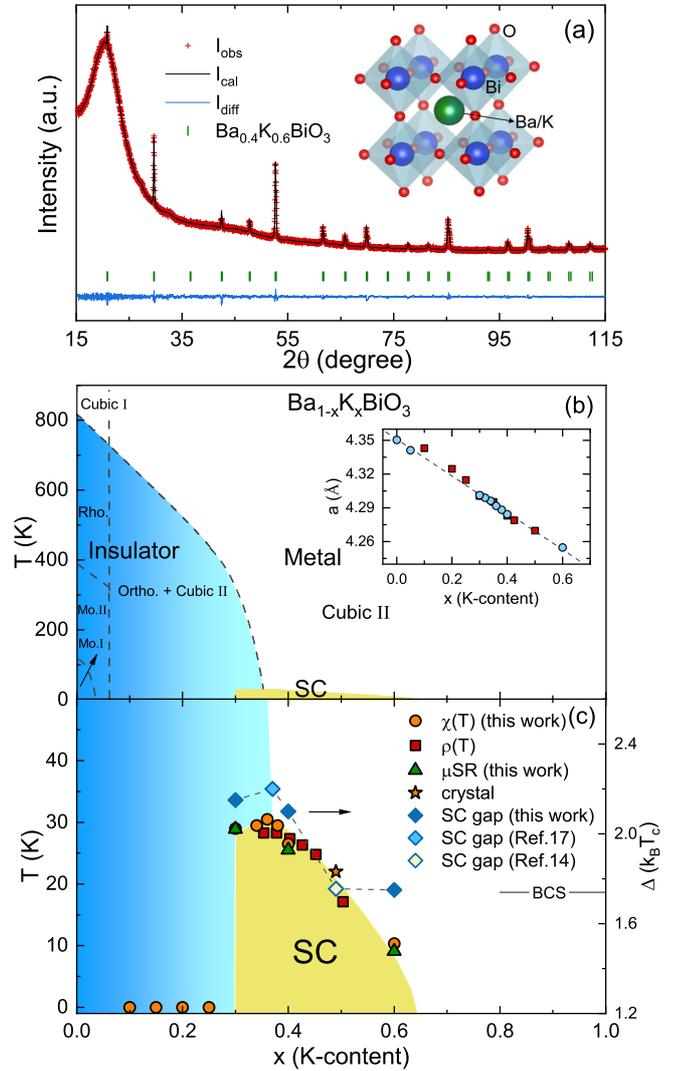}
	\vspace{-2ex}%
	\caption{\label{fig:phase_diagram}(a) A typical 
	room-temperature XRD pattern and Rietveld 
	refinement for Ba$_{0.4}$K$_{0.6}$BiO$_3$. The crystal structure 
	is shown in the inset. (b) Structural- and superconducting phase diagram of Ba$_{1-x}$K$_x$BiO$_3$, constructed according to Refs.~\onlinecite{Plumb2016,Sleight2015,Pei1990}. The inset shows the in-plane lattice parameter as a function of K-content, with the red-symbols adopted from Ref.~\onlinecite{Pei1990} and the blue symbols from our XRD data. For $0.1 \leq x \leq 0.25$, due to a mixture of phases, the relevant lattice parameter is not shown. 
    (c) Enlarged superconducting phase diagram, as obtained from  magnetization, $\mu$SR, and electrical-resistivity measurements. 
     The evolution with $x$ of the SC gap (generally scaling as the coupling strength) is shown in the right scale. The single-crystal data were taken from Ref.~\onlinecite{Wen2018}, while the electrical resistivity data from Ref.~\onlinecite{Pei1990}.
	The space groups of the different phases are: $Fm\bar{3}m$ (cubic I), $R\bar{3}$ (rhombohedral), $P2_1/n$ (monoclinic I), $I2/m$ (monoclinic II), $Ibmm$ (orthorhombic), and $Pm$-$3m$ (cubic II).} 
\end{figure}

\enlargethispage{8pt}
Polycrystalline Ba$_{1-x}$K$_{x}$BiO$_3$ 
samples were synthesized via solid-state reaction methods~\cite{Supple}.
The room-temperature XRD, measured using a Bruker D8 diffractometer, confirmed the samples' purity and the lack of 
extra phases. The atomic ratios in the various Ba$_{1-x}$K$_{x}$BiO$_3$ samples were measured by x-ray fluorescence spectroscopy (XRF) on an AMETEK Orbis Micro-XRF analyzer. The linear behavior of the in-plane lattice parameter [extracted from the Rietveld refinements -- see Fig.~\ref{fig:phase_diagram}(a) and Supplementary Materials~\cite{Supple}]
vs.\ K-content indicates the successful and homogeneous Ba/K substitution in all the studied samples [see inset in Fig.~\ref{fig:phase_diagram}(b)]. The magnetic susceptibility measurements were performed on a Quantum Design magnetic property measurement system (MPMS). The bulk $\mu$SR measurements were carried out 
at the general-purpose- (GPS), 
the multi-purpose- (Dolly), 
and the high-field and low-temperature (HAL-9500) 
surface-muon spectrometers
at the Swiss muon source of Paul Scherrer Institut, Villigen, Switzerland. 
The $\mu$SR data were analyzed by means of the \texttt{musrfit} software package~\cite{Suter2012}.   
\section{Results and discussion\label{sec:results}}\enlargethispage{8pt}
%

%

\begin{figure}[tb]
  \centering
  \includegraphics[width=0.49\textwidth]{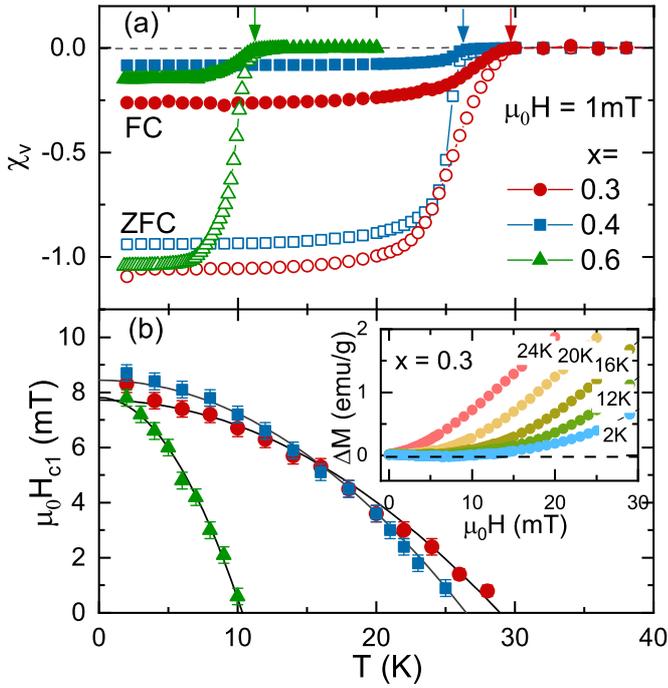}
  \caption{\label{fig:superconductivity}(a) Temperature dependent magnetic 
  susceptibility $\chi(T)$ of Ba$_{1-x}$K$_x$BiO$_3$ for $x = 0.3$, 0.4, and 0.6. 
  (b) Estimated lower critical field $\mu_{0}H_{c1}$ vs.\ temperature. 
   The solid lines are fits to $\mu_{0}H_{c1}(T) =\mu_{0}H_{c1}(0)[1-(T/T_{c})^2]$.
   Inset shows $\Delta$$M(H)$ 
   	for $x$ = 0.3 at 
   	selected temperatures; the dashed line indicating the zero value of $\Delta$$M(H)$.
   The magnetic susceptibilities were corrected by using the demagnetization
  factor obtained from the field-dependent magnetization at 2\,K (base temperature).}
\end{figure}
%
\emph{Characterization of bulk superconductivity}. The superconductivity of Ba$_{1-x}$K$_x$BiO$_3$ ($0.1 \le x \le 0.6$) was characterized by magnetic susceptibility measurements, performed in a 1-mT field, using both field-cooled (FC) and  zero-field-cooled (ZFC) protocols. As shown in Fig.~\ref{fig:superconductivity}(a), the ZFC-susceptibility, corrected to account for the demagnetization factor, indicates bulk superconductivity below $T_c$ = 29.5, 26.5, and 10.3\,K for $x$ = 0.3, 0.4, and 0.6, respectively. For $x \le 0.2$, no superconducting transition could be detected down to 1.8\,K. For $x$ = 0.25, 
the $\chi(T)$ curve shows a superconducting transition at 25\,K with a rather small superconducting fraction (below 1\%). 
The other samples, with a K-content between 0.3 and 0.4, exhibit bulk superconductivity, too (see the phase diagram). To perform 
TF-$\mu$SR measurements on superconductors, the applied magnetic field should exceed the
lower critical field $\mu_{0}H_{c1}$, so that the additional field-distribution broadening due to the flux-line lattice (FLL) can be quantified from the muon-spin relaxation rate. To determine $\mu_{0}H_{c1}$, the field-dependent 
magnetization $M(H)$ was measured at various temperatures up to $T_c$~\cite{Supple}.  
For each temperature, $\mu_{0}H_{c1}$ was determined as the value where $M(H)$ 
deviates from linearity. Here, $\Delta M(H) = M(H) - M_\mathrm{linear}(H)$, where $M_\mathrm{linear}$ is the linear low-field magnetization, as obtained from a linear fit to 
the $M(H)$ data (see Fig.~S2 in Supplementary Material). 
The inset of Fig.~\ref{fig:superconductivity}(b) shows $\Delta$$M(H)$ 
for the $x = 0.3$ case at several temperatures below $T_c$.
As indicated by the dashed line, at $\mu_{0}H_{c1}$, $\Delta M(H)$ starts 
deviating from zero value. The $M(H)$ data for all the samples 
are reported in the Supplementary Material~\cite{Supple}.  
The estimated $\mu_{0} H_{c1}(T)$ values are shown in the main panel of Fig.~\ref{fig:superconductivity}(b) as a function of temperature. The solid-lines represent fits to $\mu_{0}H_{c1}(T) = \mu_{0}H_{c1}(0)[1-(T/T_{c})^2]$ and yield lower critical fields 
of ca.\ 8\,mT for all the Bi$_{1-x}$K$_x$BiO$_3$ samples (see details in Table~\ref{tab:parameters}).

\begin{figure}[th]
	\centering
	\includegraphics[width=0.48\textwidth,angle=0]{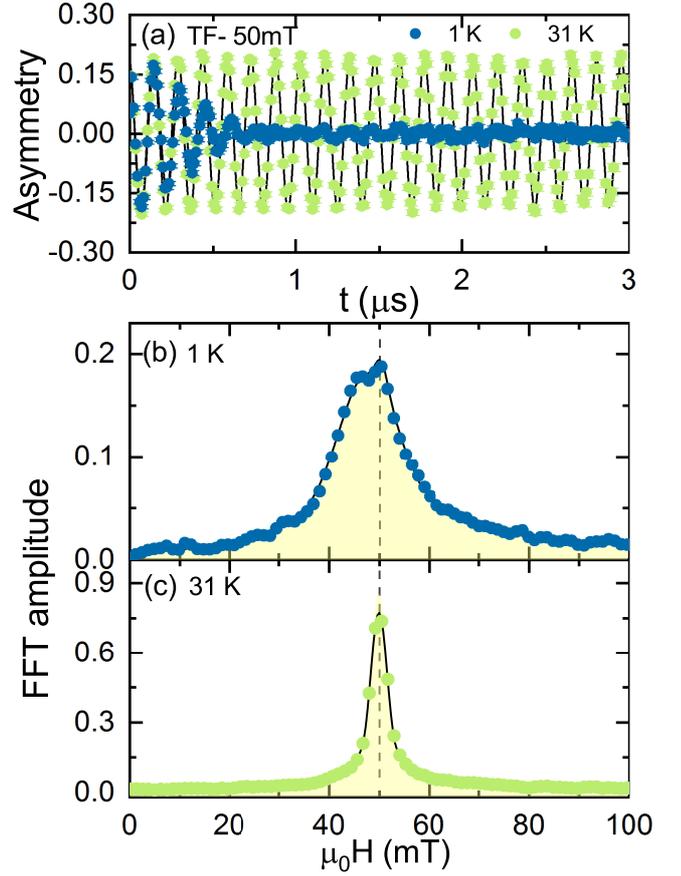}
	\vspace{-2ex}%
	\caption{\label{fig:TF_MuSR}(a) TF-$\mu$SR time-domain spectra for Ba$_{0.6}$K$_{0.4}$BiO$_3$, collected below (1\,K) and above (31\,K) $T_c$ in an applied field of 50\,mT. The other samples exhibit similar features. The fast muon-spin relaxation rate reflects the extra broadening of the field distribution due to the development of FLL.  
     Fast Fourier transforms (FFT) of the above time spectra at 1\,K (b) and 31\,K (c). The solid lines in (a)-(c) 
	are fits to Eq.~(\ref{eq:TF_muSR}) using a single Gaussian relaxation;  the dashed line indicates the applied external magnetic field (50\,mT). Note the clear diamagnetic shift below $T_{c}$ in (b).}
\end{figure}
%

\emph{Transverse-field $\mu$SR}. The TF-$\mu$SR time spectra were collected at various temperatures up to $T_c$, following a FC protocol. To track the additional field-distribution broadening due to the FLL in the mixed superconducting state, a magnetic field of 50\,mT, i.e., rather large compared to the lower critical fields of Ba$_{1-x}$K$_x$BiO$_3$, was applied above $T_c$. Figure~\ref{fig:TF_MuSR}(a) 
shows two representative TF-$\mu$SR spectra for Ba$_{0.6}$K$_{0.4}$BiO$_3$, 
collected above and below $T_c$, with the other samples 
showing similar features. The enhanced depolarization rate below $T_c$ reflects the inhomogeneous field distribution due to the FLL, causing an additional distribution broadening in the mixed superconducting state, as clearly seen from the fast-Fourier-transform (FFT) spectra shown in Fig.~\ref{fig:TF_MuSR}(b) and (c). Since the relaxations are mostly Gaussian-like, the TF-$\mu$SR-asymmetry could be modeled by: 
\begin{equation}
\label{eq:TF_muSR}
A_\mathrm{TF} = A_\mathrm{s} \cos(\gamma_{\mu} B_\mathrm{s} t + \phi) e^{- \sigma^2 t^2/2} +
A_\mathrm{bg} \cos(\gamma_{\mu} B_\mathrm{bg} t + \phi).
\end{equation}
Here $A_\mathrm{s}$ and $A_\mathrm{bg}$ represent the initial muon-spin 
asymmetries for muons implanted in the sample and sample holder (i.e., background), respectively, with the latter not undergoing any depolarization. 
$B_\mathrm{s}$ and $B_\mathrm{bg}$ are the local fields sensed by implanted muons in the sample and sample holder, $\gamma_{\mu} = 2\pi \times 135.53$\,MHz/T is the muon gyromagnetic ratio, 
$\phi$ is the shared initial phase, and $\sigma$ is a Gaussian relaxation rate. Given the nonmagnetic nature of the sample holder,  $B_\mathrm{bg}$ coincides with the applied magnetic field [see dashed line in Fig.~\ref{fig:TF_MuSR}(b)-(c)] and was used as an intrinsic reference.
%
\begin{figure}[th]
	\centering
	\includegraphics[width=0.48\textwidth,angle=0]{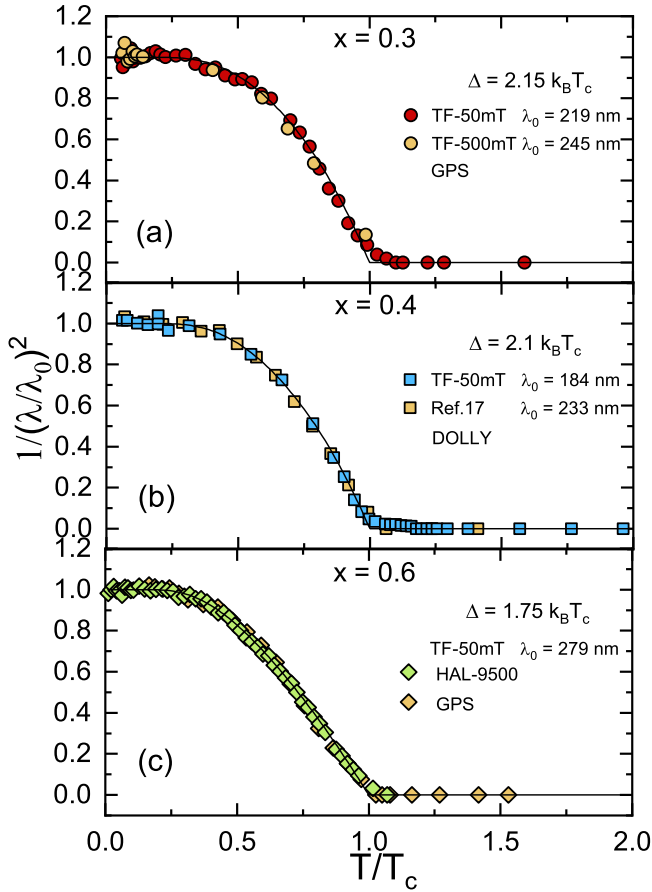}
	\vspace{-2ex}%
	\caption{\label{fig:lambda}Normalized superfluid density vs.\ temperature, as determined from TF-$\mu$SR measurements for $x$ = 0.3 (a), 0.4 (b), and 0.6 (c). 
		The solid lines represent fits to a fully-gapped $s$-wave model (see text).  
		For $x$ = 0.4, the GPS-data are highly consistent with those reported in Ref.~\onlinecite{Zhao2007} collected in a field of 200\,mT. The tiny mismatch between the theoretical value and experimental data near $T_c$ might be related to the broad superconducting transition (see Fig.~\ref{fig:superconductivity}).The fitting parameters are summarized in Table~\ref{tab:parameters}.}
\end{figure}
%

In the superconducting state, the measured Gaussian relaxation rate $\sigma$ includes contributions from both the FLL ($\sigma_\mathrm{sc}$) and a temperature-invariant relaxation due to nuclear magnetic moments ($\sigma_\mathrm{n}$) (see also ZF-$\mu$SR below). 
The FLL-related relaxation can be extracted by subtracting the nuclear contribution according to $\sigma_\mathrm{sc}$ = $\sqrt{\sigma^{2} - \sigma^{2}_\mathrm{n}}$. 
For small applied magnetic fields [with respect to the upper critical field $H_{c2}$ ($H_\mathrm{appl}$/$H_{c2}$ $\ll$\,1)], the magnetic penetration depth $\lambda$ can be obtained from $\sigma_\mathrm{sc}$$(T)$~\cite{Barford1988,Brandt2003}:
\begin{equation}
\label{eq:sig_to_lam}
\frac{\sigma_\mathrm{sc}^2(T)}{\gamma^2_{\mu}} = 0.00371\, \frac{\Phi_0^2}{\lambda^4(T)},
\end{equation}
with $\Phi_0$ the quantum of magnetic flux. For Ba$_{1-x}$K$_x$BiO$_3$, $H_{c2}$ is much higher than the applied magnetic field~\cite{Welp1988,Barilo1998}, implying the validity of the above equation. 
The derived inverse square of the magnetic penetration depths (proportional to the superfluid density $\rho_\mathrm{sc}$), are shown normalized to the 
zero-temperature values in Fig.~\ref{fig:lambda}. The temperature-independent behavior of the 
superfluid density for $T/T_c < 1/3$ clearly suggests the absence of  
excitations and, therefore, a nodeless $s$-wave 
superconductivity in Ba$_{1-x}$K$_x$BiO$_3$. 
By converse, for a nodal superconductor, the superfluid density 
is expected to depend on temperature below $T_c/3$, as e.g., 
in $p$- or $d$-wave superconductors~\cite{Bonalde2000,Khasanov2007}. 
To gain further insight into the superconducting pairing symmetry of 
Ba$_{1-x}$K$_x$BiO$_3$, the temperature-dependent superfluid density 
$\rho_\mathrm{sc}(T)$ was further analyzed by using a fully-gapped $s$-wave model:
\begin{equation}
\label{eq:rhos}
\rho_\mathrm{sc}(T) =  1 + 2\int^{\infty}_{\Delta(T)} \frac{E}{\sqrt{E^2-\Delta^2(T)}} \frac{\partial f}{\partial E} \mathrm{d}E, 
\end{equation}
where $f = (1+e^{E/k_\mathrm{B}T})^{-1}$ is the Fermi function and $\Delta(T)$ is the superconducting gap function. The temperature variation of the superconducting gap is assumed to follow 
$\Delta(T) = \Delta_0 \mathrm{tanh} \{ 1.82[1.018(T_\mathrm{c}/T-1)]^{0.51} \}$~\cite{Carrington2003}, where $\Delta_0$, the zero-temperature gap value, is the only adjustable parameter. The solid lines in Fig.~\ref{fig:lambda} are fits to a 
single-gap $s$-wave model, which yields 
magnetic penetration depths $\lambda_0$ = 219(3), 184(2), and 279(2)\,nm, 
and gap values $\Delta_0/k_\mathrm{B}T_c$ = 2.15(2), 2.10(3), and 1.75(3), 
for $x$ = 0.3, 0.4, and 0.6, respectively. 
The datasets at higher fields (500 and 200\,mT for $x = 0.3$ and 0.4, respectively) exhibit almost identical features as those at 50\,mT, further confirming the single-gap nature of superconductivity in the Ba$_{1-x}$K$_x$BiO$_3$ system. Close to optimal K-doping (e.g., $x$ = 0.3 and 0.4), the derived gap values are significantly larger than the BCS value of 1.76\,$k_\mathrm{B}T_c$; while upon over-doping (e.g., $x = 0.6$), 
the gap is more consistent with the BCS value in the weak-coupling limit. 
Since normally the SC gap scales with the coupling strength,
a progressive increase of K-doping in Ba$_{1-x}$K$_x$BiO$_3$, 
from optimal- to the over-doped regime seems to correspond 
to a change from strong- to weak electron-phonon coupling. 
While from these results one can infer that the overall coupling 
constant $\lambda_\mathrm{ep}$ decreases, this can potentially be weakened 
also by a reduction in the density of states at $E_\mathrm{F}$, rather 
than a decrease in the electron-phonon scattering matrix elements per se ~\cite{Dee2019}.
In any case, the fully-gaped state and the strong electron-phonon coupling are also supported from recent photoemission measurements~\cite{Wen2018}.  

The superfluid density and the magnetic penetration depth are
intimately linked through the London equation  $\lambda^{-2}$ =  $\mu_0 e^2 n_\mathrm{s} m^*$, where $m^*$ is the effective mass of quasiparticles and $n_\mathrm{s}$ the superfluid density~\cite{tinkham1996}. The non-monotonic dependence 
of the magnetic penetration depth in Ba$_{1-x}$K$_x$BiO$_3$ with increasing 
K-doping can be due to either a change of $m^*$ or $n_\mathrm{s}$.
With no prior knowledge of how these quantities change with doping, 
it is not straightforward to explain the observed non-monotonic 
dependence of $\lambda(x)$. 
Nevertheless, from an analogy with other unconventional superconductors, 
we can still make an educated guess. In the under-doped regime, the cuprates show a linear relationship between the superfluid density and the critical temperature~\cite{Uemura1989,Niedermayer1993}. 
On moving through the optimal doping towards the over-doped regime, this behavior 
becomes more complex and shows a ``boomerang''-like shape (whose origin 
is still under debate). 
In Ba$_{1-x}$K$_x$BiO$_3$, $T_c$ changes only slightly between $x = 0.3$ 
and 0.4, while the superfluid density increases substantially (by almost 40\%). 
On the other hand, for $x = 0.6$, both $T_c$ and the superfluid density decrease significantly. 
To conclude whether a ``boomerang''-like (i.e., non-monotonic) behavior
is confirmed also here, 
further doping values are required.
In addition, for an independent access to $m^*$ and $n_\mathrm{s}$, 
low-$T$ specific heat or Hall resistivity measurements could be very helpful.

\begin{figure}[th]
	\centering
	\includegraphics[width=0.48\textwidth,angle=0]{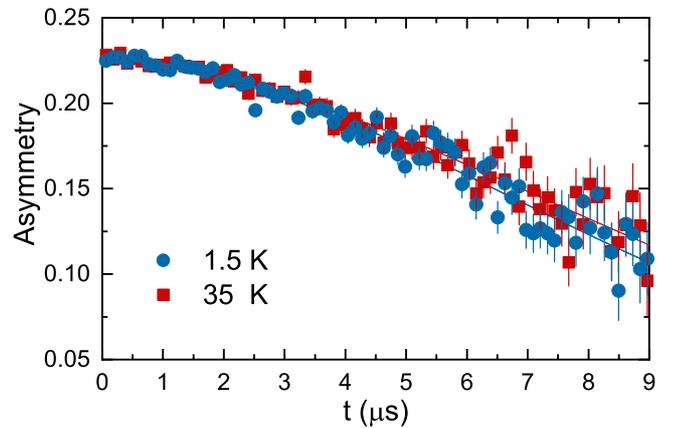}
	\vspace{-2ex}%
	\caption{\label{fig:ZF_MuSR}ZF-$\mu$SR spectra in the 
		superconducting- (1.5\,K) and the normal state (35\,K) of Ba$_{0.7}$K$_{0.3}$BiO$_3$. 
		Both data sets show only a weak muon-spin depolarization with no visible differences, implying a preserved time-reversal symmetry.
		The solid lines are fits to the spectra, as described in the text.}
\end{figure}
%

\emph{Zero-field $\mu$SR.} To search for possible magnetic features 
or time-reversal symmetry (TRS) breaking in the superconducting state 
of Ba$_{1-x}$K$_x$BiO$_3$ we performed also zero-field (ZF-) $\mu$SR measurements. 
Representative ZF-$\mu$SR spectra collected above 
and below $T_c$ for Ba$_{0.7}$K$_{0.3}$BiO$_3$ are shown in Fig.~\ref{fig:ZF_MuSR}. 
Here, the absence of coherent oscillations or fast damping is evidence of the nonmagnetic 
nature of Ba$_{1-x}$K$_x$BiO$_3$. Therefore, the weak muon-spin relaxation is mainly determined by the randomly oriented nuclear moments, which can be described by a Gaussian Kubo-Toyabe relaxation function $G_\mathrm{KT} = [\frac{1}{3} + \frac{2}{3}(1 -\sigma^{2}t^{2})\,\mathrm{e}^{-\frac{\sigma^{2}t^{2}}{2}}] $~\cite{Kubo1967,Yaouanc2011}. In polycrystalline samples, the 1/3-nonrelaxing 
and 2/3-relaxing components of the asymmetry correspond to the powder average of the local internal fields with respect to the initial muon-spin direction. The solid lines in Fig~\ref{fig:ZF_MuSR} represent fits to each dataset by considering an additional Lorentzian relaxation $\Lambda$, i.e., $A_\mathrm{ZF} = A_\mathrm{s} G_\mathrm{KT} \mathrm{e}^{-\Lambda t} + A_\mathrm{bg}$. 
Here $A_\mathrm{s}$ and $A_\mathrm{bg}$ are the same as in the TF-$\mu$SR case [see Eq.~(\ref{eq:TF_muSR})]. 
The resulting fit parameters are also summarized in Table~\ref{tab:parameters}. The weak Gaussian and 
Lorentzian relaxation rates reflect the small nuclear moments in  Ba$_{1-x}$K$_x$BiO$_3$.
In both the normal- and the superconducting states, the relaxations 
are very similar (within the experimental error), as demonstrated by 
the practically overlapping ZF-$\mu$SR spectra above and below $T_c$. 
This lack of evidence for an additional $\mu$SR relaxation below $T_c$, excludes a possible TRS breaking 
in the superconducting state of Ba$_{1-x}$K$_x$BiO$_3$. 

\begin{table}[th]
	\centering
	\caption{\label{tab:parameters}
    Superconducting parameters of 
	Ba$_{1-x}$K$_{x}$BiO$_3$ as determined from magnetization and TF-$\mu$SR and fit parameters related to ZF-$\mu$SR data collected above and below $T_c$ in Ba$_{0.7}$K$_{0.3}$BiO$_3$.} 
	\begin{ruledtabular}	
	    	Superconducting parameters vs.\ K-doping \\  
	  \begin{tabular}{lccc}
	  	Parameter                         & $x=0.3$      & $x=0.4$    & $x=0.6$     \\ \hline
	  	$T_c$($\chi$) (K)                 & 29.0         & 26.5       & 10.3        \\
	  	$T_c$($\mu$SR) (K)                & 28.9         & 25.5       & 9.1         \\
	  	$\mu_0H_{c1}$ (mT)              & 7.7          & 8.4        & 7.8         \\
	  	$\Delta$ ($k_\mathrm{B}$$T_c$)    & 2.15         & 2.10       & 1.75        \\
	  	$\Delta$ (meV)                    & 5.35         & 4.61       & 1.37        \\
	  	$\lambda_0$ (nm)\footnotemark[1]  & 219          & 184        & 279         \\
	  \end{tabular}
	  	 \vspace{8pt}
	  
	    ZF muon-spin relaxation parameters
		\begin{tabular}{lcc}
			Parameter                & at 1.5\,K                      & at 35\,K                    \\ \hline
	    	$A_\mathrm{s}$             & 0.19982(88)                 & 0.20019(91)             \\
	    	$\sigma$ ($\mu$s$^{-1}$)   & 0.0105(31)                  & 0.0120(31)              \\
	    	$\Lambda$ ($\mu$s$^{-1}$)  & 0.0963(35)                  & 0.0881(39)              \\
	    	$A_\mathrm{bg}$            & 0.02724(88)                 & 0.02729(91)             \\
	    \end{tabular}	
	\footnotetext[1]{Derived from TF-50\,mT $\mu$SR measurements.}
	\end{ruledtabular}
\end{table} 

\emph{Discussion}. In this study we pursued a twofold goal: to 
re\-li\-ably 
extend/revisit the superconducting phase diagram of 
Ba$_{1-x}$K$_x$BiO$_3$ and to reconcile
the seemingly 
contradictory mechanisms put forward to 
explain its superconductivity.

As for the first point, Ba$_{1-x}$K$_x$BiO$_3$ represents 
a very interesting system among oxide superconductors, to be compared 
against the cuprates. Prominent differences include the isotropic 
character of Ba$_{1-x}$K$_x$BiO$_3$ and its lack of half-filled $d$-orbitals, in contrast 
to the two-dimensional nature of cuprates that contain Cu in a 3$d^9$ 
state. 
However, detailed studies, as the one presented here, have been hampered 
by the lack of high-quality Ba$_{1-x}$K$_x$BiO$_3$ single crystals. 
Due to the high reactivity and volatility of K$_2$O, bulk Ba$_{1-x}$K$_x$BiO$_3$ samples 
can only be prepared in a vacuum- or dry environment 
(often at low temperatures),  
or by means of high-pressure and high-temperature techniques.  
Our successful 
systematic synthesis of Ba$_{1-x}$K$_x$BiO$_3$ 
samples with different K-doping values, 
made it possible to reliably construct the Ba$_{1-x}$K$_x$BiO$_3$ phase diagram using 
different techniques and to confirm once more (this time over the whole 
phase diagram, and not only for one composition, as in Ref.~\cite{Wen2018}), 
the \emph{conventional nature} of its superconductivity.

As for the second point, over the years, different experimental techniques 
(e.g., photoemission, tunneling, optics) have provided conflicting estimates for 
the electron-phonon coupling strength $\lambda_\mathrm{ep}$ in Ba$_{1-x}$K$_{x}$BiO$_3$. 
Such enduring controversy has been complicated by the fact that only 
rarely high-quality materials were available across the whole K-doping 
range and not always they could be studied systematically. 
By showing that the coupling regime is doping dependent, our systematic 
$\mu$SR investigation of the entire family finally clarifies this long 
standing issue and offers new insights concerning 
the pairing
mechanism in Ba$_{1-x}$K$_{x}$BiO$_3$.

\section{Summary\label{sec:summary}}\enlargethispage{8pt}
By successfully synthesizing high-quality samples of the 
Ba$_{1-x}$K$_{x}$BiO$_3$ bismuthates (with $x$ up to 0.6), we could 
systematically revisit their superconducting phase diagram.
Bulk superconductivity in the range $0.3 \le x \le 0.6$  
(with $T_c \sim 10$--30\,K) was characterized by magnetization 
measurements, followed by microscopic 
$\mu$SR experiments.
The temperature variation of the superfluid density, as determined 
via TF-$\mu$SR, reveals a fully-gapped superconductivity in Ba$_{1-x}$K$_x$BiO$_3$, 
independent of doping and well described by an isotropic $s$-wave model. 
At the same time, the derived superconducting-gap values 
strongly suggest a doping-induced \emph{crossover from strong- to weak-coupling}, 
a finding which can account 
for the seemingly contradictory models previously used to explain 
the superconductivity of the bismuthates.
Finally, the lack of spontaneous magnetic fields below $T_c$, as revealed 
by ZF-$\mu$SR measurements, indicates that time-reversal symmetry is 
preserved in the superconducting state of Ba$_{1-x}$K$_x$BiO$_3$.

\begin{acknowledgments}
This work was supported by the Schwei\-ze\-rische Na\-ti\-o\-nal\-fonds 
zur F\"{o}r\-de\-rung der Wis\-sen\-schaft\-lich\-en For\-schung, SNF 
(Grants no.\ 200021-169455 and 206021-139082). 
We thank S.\ Johnston at the University of Tennessee, Knoxville 
for useful input. 
\end{acknowledgments}

\bibliography{BKBO_bib}

\end{document}